\documentclass{aip-cp}

\usepackage[numbers]{natbib}
\usepackage{rotating}
\usepackage{graphicx}
\usepackage{adjustbox}

\def \apj{ApJ}
\def \aap{A{\&}A}
\def \araa{ARA{\&}A}
\def \pasj{PASJ}
\def \mnras{MNRAS}
\def \aj{AJ}
\def \apjl{ApJ}
\def \apss{Ap{\&}SS}
\def \ssr{SSRv}

\begin{document}

\title{Molecular Shocks and the Gamma-ray Clouds of the W28 Supernova Remnant}

\author[aff1]{Nigel Maxted\corref{cor1}}
\author[aff2]{Gavin Rowell}
\author[aff2]{Phoebe de Wilt}
\author[aff1,aff3]{Michael Burton}
\author[aff1]{Catherine Braiding}
\author[aff4]{Andrew Walsh}
\author[aff5]{Yasuo Fukui}
\author[aff6]{Akiko Kawamura}

\affil[aff1]{School of Physics, The University of New South Wales, Sydney, 2052, Australia}
\affil[aff2]{School of Physical Sciences, The University of Adelaide, Adelaide, 5005, Australia}
\affil[aff3]{Armagh Observatory and Planetarium, College Hill, Armagh, BT61 9DG, Northern Ireland, United Kingdom}
\affil[aff4]{International Centre for Radio Astronomy Research, Curtin University, GPO Box U1987, Perth, Australia}
\affil[aff5]{Department of Astrophysics, Nagoya University, Furocho, Chikusa-ku, Nagoya, Aichi, 464-8602, Japan}
\affil[aff6]{National Astronomical Observatory of Japan, Mitaka, Tokyo 181-8588, Japan}
\corresp[cor1]{Corresponding author: n.maxted@unsw.edu.au}

\maketitle
\begin{abstract}
Interstellar medium clouds in the W28 region are emitting gamma-rays and it is likely that the W28 supernova remnant is responsible, making W28 a prime candidate for the study of cosmic-ray acceleration and diffusion. Understanding the influence of both supernova remnant shocks and cosmic rays on local molecular clouds can help to identify multi-wavelength signatures of probable cosmic-ray sources. To this goal, transitions of OH, SiO, NH$_3$, HCO$^+$ and CS have complemented CO in allowing a characterisation of the chemically rich environment surrounding W28. This remnant has been an ideal test-bed for techniques that will complement arcminute-scale studies of cosmic-ray source candidates with future GeV-PeV gamma-ray observations.
\end{abstract}

\section{The W28 Region}
W28 is an old-age ($>$10$^4$\,years, see \citep{Kaspi:1993,Sawada:2012,Zhou:2014}) supernova remnant (SNR), which exhibits non-thermal synchrotron emission seen in 90\,cm radio continuum in Figure\,\ref{fig:W28}. The W28 SNR has been implicated as a cosmic ray (CR) accelerator due to a correspondence between TeV gamma-ray emission and molecular gas at a distance of 1.2-3.3\,kpc \citep{Goudis:1976,Aharonian:2008,Motogi:2011,Lozinskaya:1981}. This correspondence is true of both HESS\,J1801$-$233 - a TeV gamma-ray source immediately adjacent/overlapping the SNR, and HESS\,J1800$-$240 - a multi-component TeV gamma-ray source spatially separated by 0.25-0.75$^{\circ}$ in the plane of the sky. Such gamma-ray-gas overlap points to gamma-ray emission via the decay of neutral pions created by CR interactions with gas. It follows that the W28 SNR is a likely cosmic-ray accelerator \citep{Jones:1991,Draine:1993,Aharonian:2008}. Some uncertainty exists in the 3D geometry of the cloud complex surrounding W28 and the energy-dependent diffusion speed of CRs throughout that region. But, successive multi-wavelength campaigns have reinforced evidence for a SNR-cloud interaction and supported a hadronic scenario for gamma-ray emission.

\section{The Shocked Molecular Cloud towards HESS\,J1801$-$233}
\begin{figure}[h]
  \includegraphics[width=410pt]{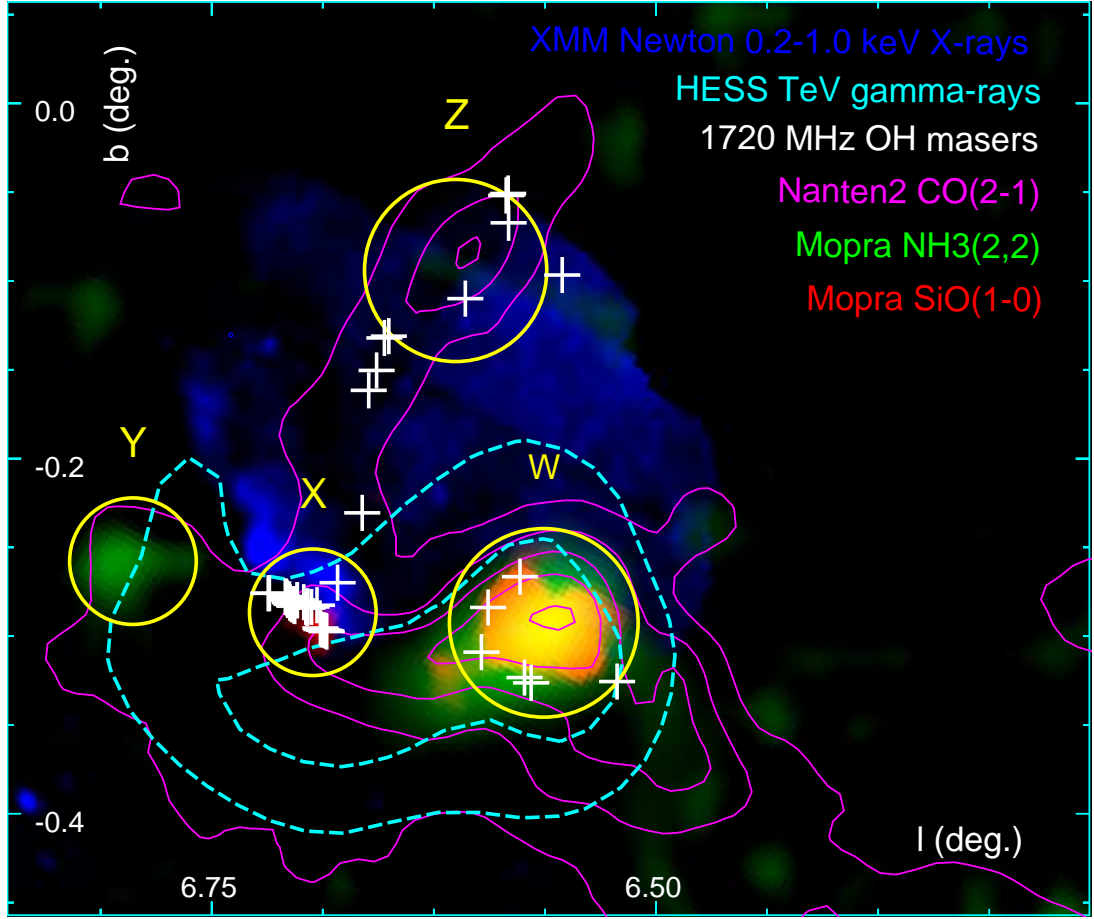}
  \caption{A 3-colour image encompassing the W28 SNR shock seen in keV X-ray (blue) emission \citep{Nakamura:2014} and the associated TeV gamma-ray source HESS\,J1801$-$233 (4, 5 and 6$\sigma$ significance contours - cyan, broken \citep{Aharonian:2008}). Mopra NH$_3$(2,2) emission (green) highlights dense, warm molecular gas clumps \citep{Maxted:2016}, while Mopra SiO(1-0) emission (red) highlights shocked molecular gas. Nanten2 CO(2-1) emission contours (magenta) show the structure of the diffuse molecular gas \citep{Fukui:2008}. White crosses indicate the locations of 1720\,MHz OH masers \citep{Frail:1994,Claussen:1997}, and yellow circles indicate regions discussed in the text (Z, Y, X and W). \label{fig:W28ne}}
\end{figure}

As illustrated in Figure\,\ref{fig:W28ne}, CO-traced molecular gas \citep{Arikawa:1999,Fukui:2008,Aharonian:2008} between kinematic velocities 0 and 20\,kms$^{-1}$ corresponds to the TeV gamma-ray emission of HESS\,J1801$-$233. A component of shock-heated molecular gas is seen in H$_2$ \citep{Neufeld:2007,Marquez:2010} and high-J CO transitions \citep{Arikawa:1999} along the radio-continuum-traced W28 shock, while 1720\,MHz OH masers (indicated in Figure\,\ref{fig:W28ne}) are also encompassed within this velocity-range \citep{Frail:1994,Claussen:1997}. 1720\,MHz OH maser lines are commonly associated with the wake of old ($\gtrsim$10$^4$\,yr) supernova remnant shocks (e.g. \citep{Claussen:1997}), and modelling suggests that very specific collisional conditions are required for their generation \citep{Elitzur:1976,Lockett:1999,Pihlstrom:2008}. C-type shocks ($\sim$25\,kms$^{-1}$) moving through dense ($\sim$10$^4$-10$^5$\,cm$^{-3}$), warm (50-125\,K) molecular gas can sustain OH maser emission provided that a source of H$_2$O-dissociation is present to maintain a large OH column density ($\sim$10$^{16}$\,cm$^{-2}$), possibly a weak UV flux emitted by H$_2$ that has been indirectly excited by CRs and/or X-ray photons \citep{Wardle:1999}. This scenario may be supported by low-level OH absorption towards W28 that indicates widespread OH abundance enhancements \citep{Pastchenko:1974,Wardle:1999}. High-J transitions of CO towards OH maser locations \citep{Arikawa:1999} support a scenario of high-temperature chemistry associated with a post-shock region. 
\begin{figure}[h]
  \centerline{\includegraphics[width=410pt]{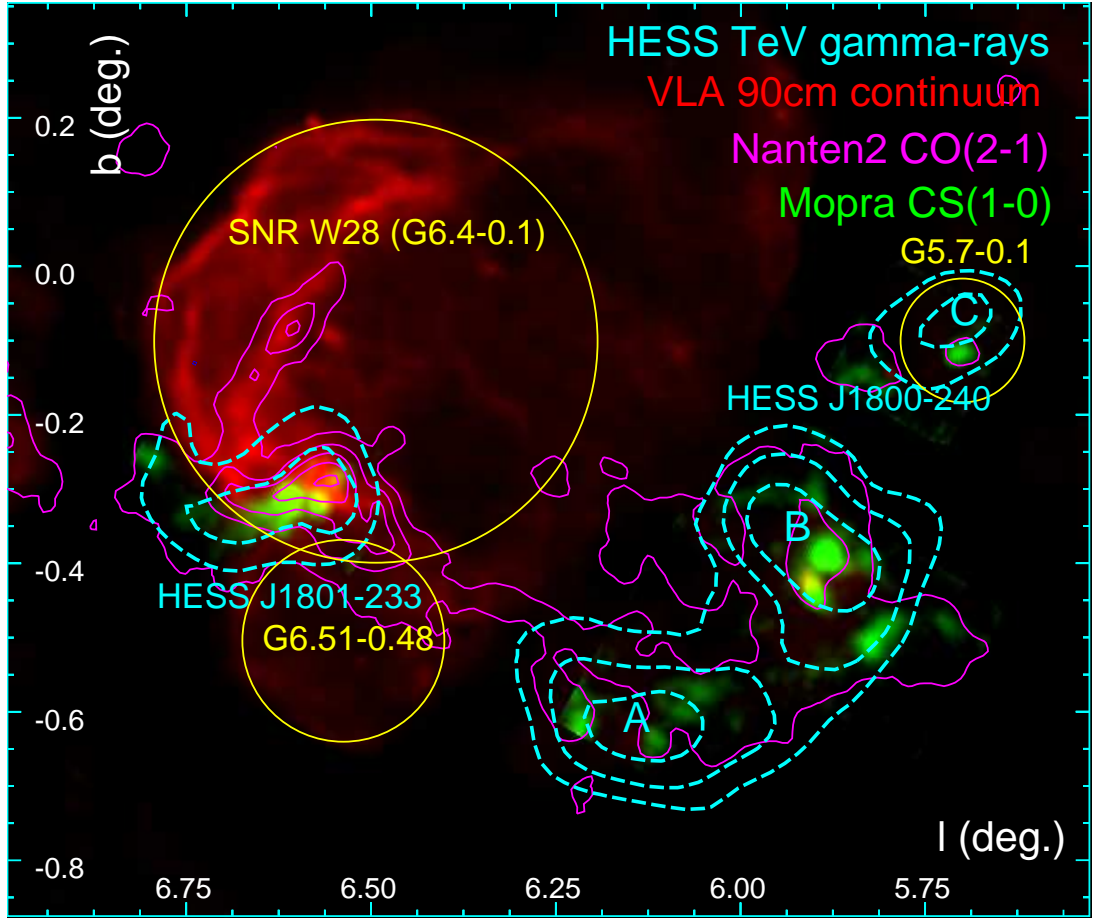}}
  \caption{The W28 supernova remnant, as seen in 90\,cm radio continuum \citep{Dubner:2000} emission (red), and associated dense gas clumps highlighted by velocity-integrated ($-$5 to 20\,kms$^{-1}$) Mopra CS(1-0) emission (green) \citep{Nicholas:2011}. HESS TeV gamma-ray emission 4, 5 and 6$\sigma$ significance contours \citep{Aharonian:2008} (cyan, broken) and velocity-integrated ($-$10 to 25\,kms$^{-1}$) Nanten2 CO(2-1) emission \citep{Fukui:2008} contours are overlaid (magenta). The approximate locations of the W28, G6.51$-$0.48 and G5.7$-$0.1 SNRs are indicated by yellow circles, while HESS gamma-ray sources HESS\,J1801$-$233 and HESS\,J1800$-$240 (A, B and C) are indicated. \label{fig:W28}}
\end{figure}

Given the geometry of supernova shock-cloud interactions with respect to the observer, velocity-coherent OH column density is maximized perpendicular to shock motion, therefore the line-of-sight velocities of OH masers are generally expected to be representative of the systemic velocity of gas associated with the object that injected the shock - in this case the W28 SNR.
OH masers towards W28 have line-of-sight velocities ranging between $\sim$5 and $\sim$16\,kms$^{-1}$. Given the proximity to the Galactic Centre, there is some confusion in distinguishing which Galactic arm the masers and the W28 gas correspond to. The kinematic velocity range of the gas is consistent with both the Sagittarius ($\sim$0-2.5\,kpc) and Scutum-Crux ($\sim$2.5-4\,kpc) arms \citep{Aharonian:2008}.

The C-shocks of W28 also lead to the production of gas-phase SiO in post-shock regions. \citet{Nicholas:2012} mapped the $\sim$5$\times$10$^4$\,M$_{\odot}$-mass of dense gas clumps within the HESS\,J1801$-$233 cloud in CS(1-0) emission (Figure\,\ref{fig:W28}), while simultaneously observing SiO(1-0) emission along the boundary of the W28 shock (Regions X and W, highlighted in Figure\,\ref{fig:W28ne}). As observed towards other shocked regions (e.g. IC\,443 \citep{Ziurys:1989_ic443,Turner:1992,vanDishoeck:1993} and the Galactic Centre \citep{Martin-Pintado:2000,Minh:2015}), silicon is liberated from dust grain mantles in regions where $\gtrsim$25\,kms$^{-1}$-velocity shocks propagate through dense ($\sim$10$^{5}$\,cm$^{-3}$) gas \citep{Schilke:1997,May:2000,Gusdorf:2008}, while SiO production is driven in the high-temperature ($\sim$100\,K) wake \citep{Ziurys:1989_HighT,Langer:1990} through chemical pathways that include the destruction of OH molecules \citep{Hartquist:1980,Langer:1990}. A weight of evidence thus supports the notion that SiO emission is an unambiguous tracer of shocks, while near W28, positional coincidence with other tracers reinforces the view that SiO is tracing dense, post-SNR-shocked molecular gas.
%


Evidence of a shock interaction are also present in NH$_3$ emission surveys \citep{Nicholas:2011,Maxted:2016}. Like CS, NH$_3$ probes the densest components of the CO-traced molecular clouds (see Figure\,2), but the hyperfine structure and multiple energetic states of the NH$_3$ inversion transition allow direct optical depth and temperature measurements with decreased systematic uncertainty. Energetic NH$_3$(n,n) emission (where n=3,4,6) and an elevated velocity dispersion are seen towards the region exhibiting the strongest SiO emission (Region\,W of Figure\,\ref{fig:W28ne}) on the W28 SNR-side of cloud, reinforcing the scenario that W28 is a source of cloud-disruption \citep{Maxted:2016}. Furthermore, signatures of dust grain sputtering may be present in the elevated abundance of ortho-NH$_3$ relative to para-NH$_3$ \citep{Maxted:2016}. Ground rotational state ortho-NH$_3$ molecules require less energy to separate from dust grain surfaces, so are liberated more easily in shocks than ground rotational state para-NH$_3$ molecules leading to a disequilibrium \citep{Umemoto:1999}. Once such an imbalance is occurs, a population of NH$_3$ molecules can take $\sim$10$^6$\,yr to return to statistical equilibrium via forbidden rotational transitions \citep{Cheung:1969}.

In addition to NH$_3$, CO, CS, OH and SiO detections, 36 and 44\,GHz CH$_3$OH maser emission was also detected towards a subregion of Region\,W of Figure\,\ref{fig:W28ne} \citep{Nicholas:2012,Pihlstrom:2014}. So-called `class I' CH$_3$OH masers are commonly associated with the molecular shocks of HII regions, but in the case of Region\,W, no such counterpart was identified, leaving the W28 shock as the most plausible trigger mechanism \citep{Nicholas:2012}. CH$_3$OH masers might be  a more-numerous, but less-intense, tracer of SNR shocks than 1720\,MHz OH masers \citep{Pihlstrom:2014}. Furthermore, the existence of cospatial CH$_3$OH masers and NH$_3$(3,3) emission may indicate SNR-shocked dense clumps \citep{McEwen:2016}, consistent with current single-dish observations of W28, but not yet shown on the small scales ($\sim$0.2$^{\prime}$) suggested by NH$_3$ observations \citep{Maxted:2016}.

Class\,I 44\,GHz CH$_3$OH masers are also detected coincident with the boundary of HESS\,J1801$-$233 (towards Region\,Y of Figure\,\ref{fig:W28ne}), and are associated with a dense molecular clump, hereafter referred to as the ``Region\,Y clump'', at a line of sight central velocity of $\sim$21\,kms$^{-1}$, which is $\sim$8-12\,kms$^{-1}$ (see ``Core\,1'' in \citep{Nicholas:2011}) faster than the central velocity of the gas component unambiguously shown to be associated with W28 and HESS\,J801$-$233. The CH$_3$OH masers of the Region\,Y clump cannot, by themselves, be used as evidence of a passing SNR-shock or for an association with W28-shocked gas, due to the possible association of a HII region (IRAS\,17589$-$2312 \citep{Faundez:2004}), but proof of a W28-association with the Region\,Y clump may be reflected in the DCO$^+$/HCO$^+$ ratio.

The estimated DCO$^+$/HCO$^+$ ratio ($\sim$0.003-0.02) implies an ionization rate $\sim$100$\times$ larger than in quiescent regions, possibly a consequence of 0.1-1.0\,GeV CRs accelerated in the W28 SNR shock \citep{Vaupre:2014}. It is unclear if photo-ionization from objects associated with the coincident infrared point source IRAS\,1758.9$-$2312 contributes to the measured ionization rate, but this may be unlikely given the spatial separation between several HCO$^+$ measurements within Region\,Y. Indeed, a key assumption of this analysis is that the sampled gas is in small-scale clumps which are shielded from UV radiation, but are penetrable by ionizing CRs.

Interestingly, the estimated ionization rate of $\sim$10$^{-15}$\,s$^{-1}$ for the Region\,Y clump is the same as that required to sustain an adequate abundance for the observed shock-tracing 1720\,MHz OH maser emission \citep{Wardle:1999}. Unfortunately, the DCO$^+$/HCO$^+$ ratio was not estimated in gas components directly shocked by the SNR due to sensitivity limits.


\section{W28 and HESS\,J1800$-$240}
HESS\,J1801-233 has been clearly shown, via numerous molecular shock tracers, to be associated with dense gas clumps at line of sight velocities ranging between 7 and 13\,kms$^{-1}$. A correlation between gamma-rays and the gas clearly points to a hadronic origin for the gamma-ray emission, implicating W28 as an accelerator of TeV CRs, while constraining the W28 kinematic distance to be approximately that of the shocked cloud. As stated above, at the boundary of HESS\,J1801$-$233 is a gas clump at $\sim$21\,kms$^{-1}$ that has been shown to be experiencing an ionization rate $\sim$100$\times$ larger than quiescent regions, suggesting that an ionizing flux of 0.1-1.0\,GeV CRs accelerated locally in the W28 shock pervades this gas component. The range of line-of-sight velocities of gas associated with W28 and HESS\,J1801$-$233 (7-21\,kms$^{-1}$) encompasses the line of sight velocities of nearby molecular clumps, which correspond to HESS\,J1800$-$240 (e.g. see Figure\,2 of \citep{Nicholas:2012}), thus the line of sight velocities of HESS\,J1800$-$240 and HESS\,J1801$-$233 are consistent with a similar distance for both sources. Furthermore, in Figure\,\ref{fig:W28} a bridge of CO emission can be seen to connect the 2 largest HESS gamma-ray sources, further suggesting an association. 

Not all parallax distance measurements of objects thought to be associated with HESS\,J1800$-$240 are in agreement. Distance estimates range from $\sim$1.3\,kpc \citep{Motogi:2011} to $\sim$3\,kpc \citep{Sato:2014} leading to some confusion. A larger sample of parallax measurements in this region would help to minimize distance uncertainties. Certainly though, a separate CR source (W28) would need to be invoked to explain HESS\,J1800$-$240 if it was not at the same distance as W28. HII regions and active star formation within HESS\,J1800$-$240\,A and B cannot be strictly ruled out as CR sources (e.g. see \citep{Hampton:2016} and references therein), but W28 is already confidently shown to be a likely CR source illuminating the nearby shocked cloud (HESS\,J1801$-$233), making it the prime candidate for gamma-ray illumination of HESS\,J1800$-$240\,A and B. 

HESS\,J1800$-$240\,C, on the other hand, (see Figure\,\ref{fig:W28}) may reasonably originate from an unassociated complex, as evidenced by a positional coincidence with another CR acceleration candidate SNR G5.7$-$0.1 and OH maser emission \citep{Joubert:2016}. We note that G6.51$-$0.48 (see Figure\,\ref{fig:W28}) has no strong molecular cloud-interaction signatures, and is not currently considered a strong CR-acceleration candidate.

\begin{figure}[h]
  \centerline{\includegraphics[width=425pt]{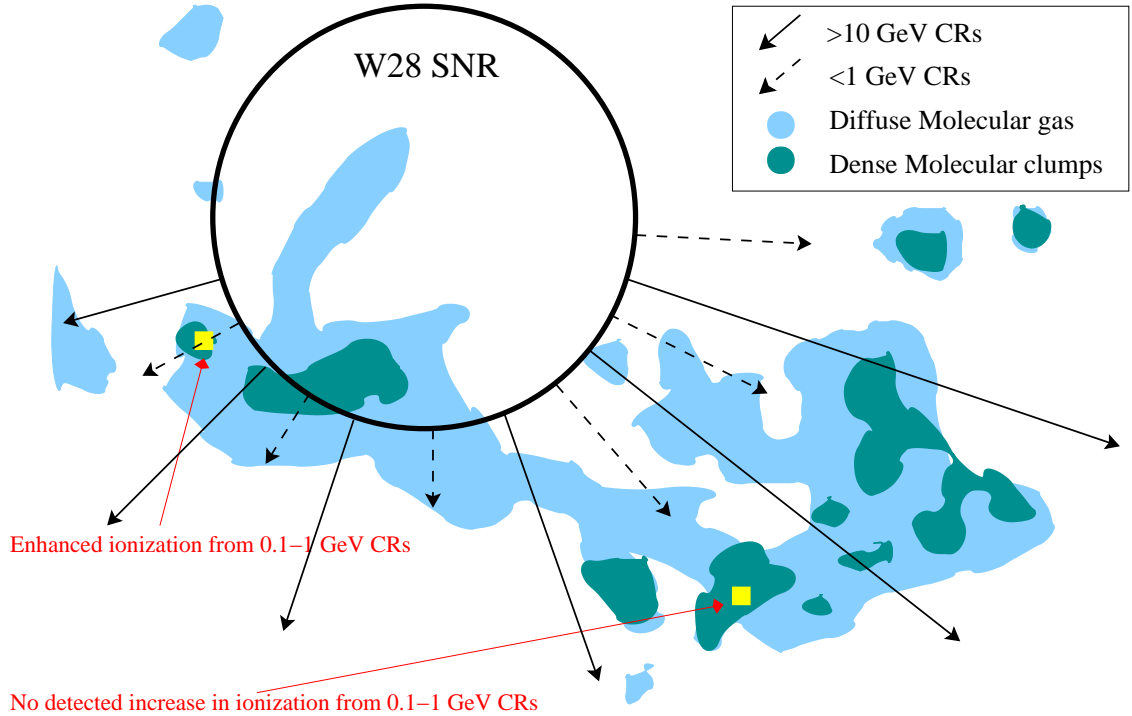}}
  \caption{A schematic of a possible scenario occurring towards the W28 region. The W28 SNR interacts with molecular gas and accelerates CR protons, which interact with gas via p-p collisions to produce the observed gamma-ray emission. The highest energy CRs preferentially escape the SNR first and reach nearby molecular gas, while lower energy ionizing CRs remain confined for longer. The resultant gamma-ray emission is largest towards the densest clumps. This scenario assumes one source of CRs in the region.\label{fig:W28schematic}}
\end{figure}

Some gas inferred to be associated with W28 does not display gamma-ray emission, possibly due to sensitivity limits of HESS and resolution limits of Fermi-LAT. 1720\,MHz OH masers in Region\,Z in Figure\,\ref{fig:W28ne} are towards a significant CO filament which extends across the SNR. This gas exhibits correspondence between synchrotron radiation and shocked H$_2$ \citep{Reach:2005}, but, like the ionized gas in Region\,Y, Region\,Z has no clear TeV gamma-ray counterpart. HESS\,J1801$-$233 is a $\sim$6$\sigma$ detection that is produced from a CR target mass of $\sim$10$^{4}$\,M$_{\odot}$, so assuming a TeV CR density similar to that of HESS\,J1801$-$233, and linear proportionality between gamma-ray flux and gas mass, a component of a few hundred solar masses is unlikely to contribute significantly to the total TeV gamma-ray flux. The Region\,Y dense clump mass is $\sim$50-140\,M$_{\odot}$ (assuming a 2\,kpc distance) \citep{Nicholas:2011}, while the Region\,Z gas had no clear sign of dense clumps in sensitive NH$_3$ emission maps \citep{Maxted:2016}, which places a dense gas mass upper limit of a few$\times$10$^2$-10$^3$\,M$_{\odot}$ (assuming a 2\,kpc distance and [NH$_3$]/[H$_2$]$\sim$10$^{-8}$-10$^{-9}$). The Region\,Z and Y clumps are thus not expected to be above current gamma-ray detection sensitivities.

A scenario incorporating CRs accelerated by the W28 shock escaping into surrounding molecular clouds can plausibly explain HESS\,J1801-233 and HESS\,J1800$-$240, as illustrated in Figure\,\ref{fig:W28schematic}. In this scenario, GeV to TeV-energy CRs penetrate into both the shocked cloud (HESS\,J1801-233) and the nearby clouds (HESS\,J1800$-$240\,A and B) to generate the observed gamma-ray emission via neutral pion creation and decay \citep{Aharonian:2008,Abdo:2010,Hanabata:2014}. The release of CRs from the SNR shell and speed of diffusion is expected to be energy-dependent, with the highest energy CRs reaching the nearby clouds before lower energies. Modelling based on the high energy spectra of HESS\,J1800$-$240 (A and B) and HESS\,J1801-233 has suggested that CR diffusion in the W28 region is slower than the Galactic average (see e.g. \citep{Gabici:2007,Gabici:2010,Nava:2013,Hanabata:2014} for details). Ionization rate estimates towards HESS\,J1800$-$240\,B indicate that the density of 0.1-1\,GeV CRs is consistent with Galactic norms \citep{Vaupre:2014}, but uncertainties in CR ionization calculations are also consistent with a slight ($\lesssim$10$\times$) ionization rate increase. Given the energy-dependent nature of CR escape from SNRs and transport over a sky-plane-projected distance of $\sim$10\,pc, a non-detection of 0.1-1\,GeV CRs is not inconsistent with predictions, and will likely be a strong constraint in future diffusion modelling. 

\section{Conclusion}
Molecular species such as OH, SiO, NH$_3$, HCO$^+$ and CS have complemented CO, and play a crucial role in probing gas that emits gamma-rays via p-p collisions, gas that has chemical signatures triggered by passing shocks, and clumps penetrated by ionizing CR radiation. With the upcoming Cherenkov Telescope Array offering unprecedented gamma-ray sensitivity and arcminute-scale angular resolution \citep{Acharya:2013}, observations of these tracers using existing millimetre-instruments will complement studies of CR origin by offering angular resolutions similar to (or better than) next generation gamma-ray data.

\bibliographystyle{aipnum-cp}%
\bibliography{ReferencesW28}

\end{document}